\documentclass[11pt]{iopart}
\usepackage{hyperref}
\usepackage{amssymb}
\usepackage{amsopn}

\usepackage{subfigure}
\usepackage{graphicx}
\usepackage{bm}

\newcommand{\be}{\begin{equation}}
  \newcommand{\ee}{\end{equation}}
\newcommand{\ba}{\begin{eqnarray}}
  \newcommand{\ea}{\end{eqnarray}}

\renewcommand{\(}{\left(}
\renewcommand{\)}{\right)}
\renewcommand{\[}{\left[} 
\renewcommand{\]}{\right]}

  \newcommand{\h}{{\cal H}}

  \newcommand{\de}{\delta}
  \newcommand{\De}{\Delta}
  \newcommand{\al}{\alpha}
  \newcommand{\Om}{\Omega}
  \newcommand{\rd}{{\rm d}}
  \newcommand{\tw}{\tilde{w}_d}
  \newcommand{\Tc}{{\rm Tc}}

\begin{document}

\title{Non-adiabatic perturbations in Ricci dark energy model}

\author{Khamphee Karwan$^{1,2}$ and Thiti Thitapura$^1$}

\address{$^1$
Department of Physics,\\ Faculty of Science, Kasetsart University, Bangkok 10900, Thailand \\
$^2$Thailand Center of Excellence in Physics,\\ CHE, Ministry of Education, Bangkok 10400, Thailand
}

\begin{abstract}

We show that the non-adiabatic perturbations between Ricci dark energy and matter can grow both on superhorizon and subhorizon scales,
and these non-adiabatic perturbations on subhorizon scales can lead to instability in this dark energy model.
The rapidly growing non-adiabatic modes on subhorizon scales always occur when the equation of state parameter of dark energy starts to drop towards $-1$
near the end of matter era, except that the parameter $\al$ of Ricci dark energy equals to $1/2$.
In the case where $\al = 1/2$, the rapidly growing non-adiabatic modes disappear when the perturbations in dark energy and matter are adiabatic initially.
However, an adiabaticity between dark energy and matter perturbations at early time implies a non-adiabaticity between matter and radiation, this can influence the ordinary Sachs-Wolfe (OSW) effect.
Since the amount of Ricci dark energy is not small during matter domination, the integrated Sachs-Wolfe (ISW) effect
is greatly modified by density perturbations of dark energy, leading to a wrong shape of CMB power spectrum.
The instability in Ricci dark energy is difficult to be alleviated if the effects of coupling between baryon and photon on dark energy perturbations are included.

\vspace{3mm}
\begin{flushleft}
CMBR
\end{flushleft}

\end{abstract}

\maketitle

\section{Introduction}

The present acceleration of the universe, suggested by various observations \cite{sn, cmb, sdss},is one of the most important puzzles in cosmology.
Possible explanations for this puzzle can be based on the modification of theory of gravity or introducing a new form of energy
whose pressure is negative called dark energy \cite{Frieman:08} - \cite{Caldwell:98}.
Many models of dark energy have been proposed in literature from various motivations.
The interesting one is motivated by the holographic principle \cite{Cohen:98}.

The holographic principle implies a relation between
the short distance (UV) cut-off and the long distance (IR) cut-off in quantum field theory \cite{Cohen:98}.
Thus the amplitude of vacuum energy, which may play the role of dark energy,
is controled by the long distance cut-off whose length scale should be comparable with one of the cosmological length scales.
Choosing the Hubble radius as an IR cut-off, the present amplitude of vacuum energy is in agreement with observations \cite{Cohen:98}.
Nevertheless, the vacuum energy evolves like matter, so that it cannot play the role of dark energy.
The equation of state parameter of vacuum energy can be negative if
 particle or future event horizons are used as the IR cut-off \cite{Fischler:98, Li:04},
but for the choice of particle horizon it cannot be negative enough to drive the present acceleration of the universe.
So we are left with the future event horizon but this choice encounters a causality problem \cite{Cai:07}.
 It has been shown that with the inverse square root of the Ricci scalar, which is propertional to the curvature radius of spacetime, as  the IR cut-off,
vacuum energy can be a viable dark energy candidate,  the Ricci dark energy \cite{Gao:07} where
\be
\rho_d = \frac \al 2 R\,,
\label{rc}
\ee
where $\rho_d$ is the energy density of dark energy, $R$ is the Ricci scalar
and $\al$ is a constant parameter. The reduced Planck mass, $M_p = 1/ \sqrt{8\pi G}$, is set to be unity.

In order to study the dynamics of cosmological perturbations for Ricci dark energy, we follow the literature \cite{Feng:09}
to suppose that the above equation is also valid in the perturbed spacetime,
which should be possible because the relation (\ref{rc}) can also be obtained from
some arguments about quantum gravity effects \cite{Gao:07}.
In \cite{Feng:09}, the adiabaticity between dark energy and matter perturbations is assumed, i.e. $\de\rho_m / \de\rho_d = \dot\rho_m / \dot\rho_d$,
where $\de\rho_d$, $\rho_m$ and $\de\rho_m$ is the density perturbations in dark energy, energy density of matter and density perturbations in matter respectively.
For this case, the instability can occur in the future, and can be avoided if $\al > 1/3$.
In this work, we will check whether the adiabaticity between dark energy and matter is preserved in general,
and study the dynamics of non-adiabatic perturbations between dark energy and matter as well as the instability in Ricci dark energy model.
We suppose that the dynamics of Ricci dark energy can be presented in terms of the perfect fluid and the conservation of energy-momentum.

In section 2, we derive some useful relations for the background dynamics of Ricci dark energy.
The dynamics of density perturbations is studied in section 3, and the conclusions are given in section 4.

\section{Background dynamics}

Using the trace of Einstein equation, we can write the energy density and pressure in terms of Ricci scalar as
\be
R = \rho_m + \rho_d - 3 p_d\,.
\label{tr1}
\ee
Here, $p_d$ is the pressure of dark energy and $\rho_m = \rho_c + \rho_b$,
where $\rho_c$ and $\rho_b$ is the energy density of Cold Dark Matter (CDM) and baryon respectively.
Substituting (\ref{rc}) into the above equation, we get
\be
p_d = \frac 13 \[\rho_m + \rho_d\(1 - \frac 2\al\)\]\,,
\ee
so that the equation of state parameter of dark energy is given by
\be
w_d = \frac 13\(r + 1 - \frac 2\al\)\,,
\label{wd}
\ee
where $r = \rho_m / \rho_d$.
The evolution equation for $r$ can be obtained from the energy conservation of matter and dark energy
\be
\dot r = \frac{d r}{d t} = \frac{\rho_m}{\rho_d}\(\frac{\dot\rho_m}{\rho_m} - \frac{\dot\rho_d}{\rho_d}\) = 3 H w_d r\,,
\label{dotr}
\ee
where $H = \dot a / a$ is the Hubble parameter and $a$ is the cosmic scale factor.
Using (\ref{wd}), the solution of the above differential equation is
\be
r = \frac{\tw A a^{\tw}}{1 - A a^{\tw}}\,.
\label{r}
\ee
Here, $\tw = 1 - 2 / \al$ and
\be
A = \frac{r_0}{r_0 + \tw}\,,
\ee
where $r_0$ is the present value of $r$. We will use subscript ${}_0$ to denote the present value of the variables through out this paper.
Since $\rho_m = \rho_{m0} a^{-3}$, the evolution of $\rho_d$ can be presented in terms of $r$ as
\be
\rho_d = \frac{\rho_{m0}}{r} a^{-3}\,.
\ee
In order to obtain the present acceleration of the universe, $\tw$ must be negative, so that in the early epoch $r$ is nearly constant and given by
\be
r_{\rm early} = - \tw = \frac 2\al - 1\,.
\label{ri}
\ee
Thus, it follows from (\ref{wd}) that dark energy evolves like a matter in the early epoch, i.e., $w_d \approx 0$.

\begin{figure}[ht]
\includegraphics[height=0.4\textwidth, width=0.9\textwidth,angle=0]{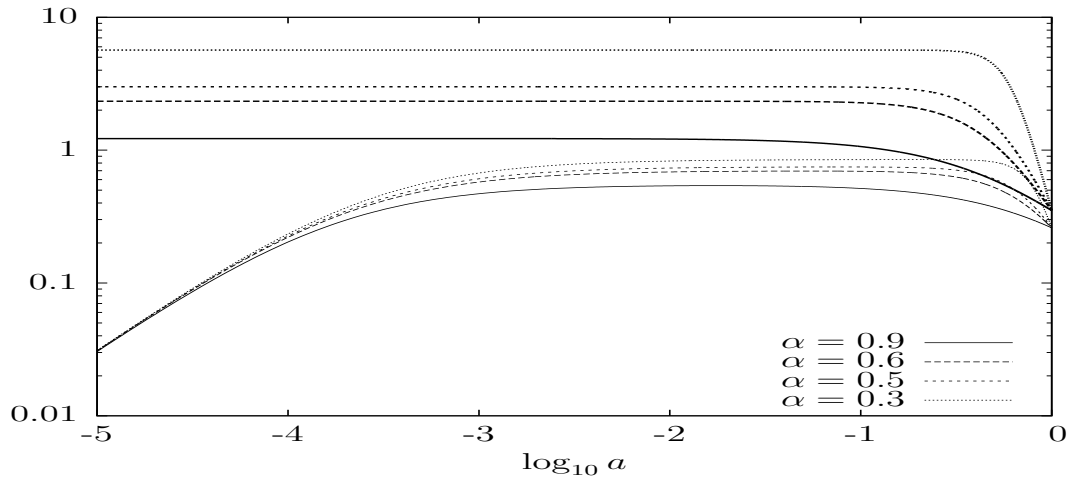}
\caption{
The evolution of $r$ for various value of $\al$ is presented by thick lines,
while the corresponding $\Om_m$ is presented by thin lines.
The present values of $\Om_m$ and $\Om_d$ are chosen such that $\Om_{m0} = 0.26$ and $\Om_{d0} = 0.74$ respectively.
}
\label{fig:1}
\end{figure}

In figure (\ref{fig:1}), the evolutions of $r$ for various values of $\al$ are plotted. From the figure we see that
when looking back in time $r$ reaches $r_{\rm early}$ rather quickly, i.e. (\ref{ri}) is approximately  valid during matter domination.
Thus, during matter domination, the density parameter of matter and dark energy is given by
\be
\Om_m \simeq 1 - \frac \al 2\,,
\qquad\qquad
\Om_d = \frac \al 2\,.
\label{om-md}
\ee
Since $\al$ should be in the range $1/3 < \al < 1$ \cite{Feng:09}, (\ref{om-md}) shows that $\Om_m$ is smaller than unity and $\Om_d$ is not small during matter domination.
We will see in the next section that this feature greatly influences the ISW effect.

%

\section{Evolution of perturbations}

In order to study dynamics of density perturbations, we start by perturbing the trace of Einstein equation around FRW background as
\be
\de R = \de \rho_m + \de \rho_d - 3\de p_d\,.
\label{dr}
\ee
Since we suppose that (\ref{rc}) is also valid in the perturbed spacetime,
the above equation can be written as
\be
\de p_d = \frac 13 \[\de\rho_m + \de\rho_d\(1 - \frac 2\al\)\]\,.
\label{dp}
\ee

Unlike the background case, the perturbations in CDM and baryon evolve differently,
so that we have to analyze their effects on the perturbations in dark energy separately.
We will express all perturbed quantities in Newtonian gauge in which the line element is
\be
\rd s^2 = - (1 + 2\Psi) \rd t^2 + a^2(t)(1 - 2\Phi)\delta_{ij}\rd x^i \rd x^j\,.
\label{ds}
\ee
According to the conservation of energy momentum tensor,
the evolutions of density contrast $\de_c = \de\rho_c / \rho_c$ and velocity perturbations $v_c$ of CDM obey \cite{Malik:08, Doran:03}
\ba
\de_c' &=& 3\Phi' - k v_c\,,
\label{dec}\\
v_c' &=& - \h v_c + k \Psi\,,
\label{vc}
\ea
while the evolution equations for density contrast $\de_b$ and velocity perturbation $v_b$ of baryon are given by \cite{Ma:95, Doran:05}
\ba
\de_b' &=& 3\Phi' - k v_b\,,
\label{deb}\\
v_b' &=& - \h v_b + k \Psi  + c_s^2 k \de_b + R\tau_c^{-1}\(v_\gamma - v_b\)
= - \h v_b + k \Psi + \Tc\,,
\label{vb}
\ea
where a prime denotes derivative with respect to conformal time $\eta = \int dt / a$,
a subscript $\gamma$ denotes photon,
$c_s^2$ is the sound speed of baryon, 
$R \equiv (4/3) \rho_\gamma / \rho_b$ and
$\tau_c$ is the Compton mean free path of a photon scattering off a baryon.
Here, $\Tc$ depends on the baryon sound speed and the coupling between baryon and photon.
In the early epoch, photon and baryon are tightly coupled, so that \cite{Doran:05}
\ba
\fl
\Tc &=& \frac{1}{1+R}k c_s^2 \de_b
+\frac{R}{1+R}\[k\(\frac 14 \de_\gamma - \sigma_\gamma \) + \h v_b+ \dot{\cal V}\]\,,
\\
\fl
\dot{\cal V} &\equiv & \Bigg \{ \[\frac{\tau'_c}{\tau_c} - \frac{2}{1+R}\]\h\(v_b - v_\gamma\) + \frac{\tau_c}{1+R} \bigg[  -\frac{a''}{a} v_b 
+k \( \frac 12 \de_\gamma - 2\sigma_\gamma + \Psi\)
\nonumber \\
\fl
&& + k\(c_s^2\de'_b - \frac 14 \de'_\gamma + \sigma'_\gamma\) \bigg] \Bigg\} 
\Bigg / \left\{1 + 2 \h \frac{\tau_c}{1+R}\right\}\,,
\ea
where $\sigma_\gamma$ is the anisotropic perturbations in photon.

The evolution equations for the perturbations in dark energy can also be obtained from
the conservation of energy momentum tensor.
In the case where the anisotropic perturbations in dark energy is absent, one gets \cite{Malik:08, Doran:03}
\ba
\de_d' &=& - 3\h\(\frac{\de p_d}{\de\rho_d} - w_d\)\de_d + 3\(1+w_d\)\Phi' - k u_d\,,
\label{dot-del}\\
u_d' &=& \h\(3w_d - 1\) u_d + k \frac{\de p_d}{\rho_d} + (1+w_d) k \Psi\,,
\label{dot-u}
\ea
where $u_d = (1 + w_d) v_d$ is the momentum density of dark energy.

Substituting (\ref{dp}) into the above equations yields
\ba
\de_d' &=& - \h r \(\de_m - \de_d\) + 3\(1+w_d\)\Phi' - k u_d\,,
\label{ded}\\
u_d' &=& \h\(3w_d - 1\) u_d + k \[\frac r3 \(\de_m - \de_d\) + w_d \de_d\] + (1+w_d) k \Psi\,,
\label{ud}
\ea
where $\de_m = (\rho_c\de_c + \rho_b\de_b) / \rho_m$.
These equations can be solved if the evolution of $\de_m$ is known.
Thus, we use (\ref{dec}) - (\ref{vb}) to derive the evolution
equation for $\de_m$ as
\ba
\de_m' &=& 3\Phi' - k v_m\,,
\label{dem}\\
v_m' &=& - \h v_m + k \Psi + \frac{\rho_b}{\rho_m}\Tc\,.
\label{vm}
\ea
Here, $v_m = (\rho_c v_c + \rho_b v_b) / \rho_m$.

In our case, it is more interesting to describe the perturbations in dark energy in terms of the following variables
\be
\De = (1+w_d)\de_m - \de_d\,,
\qquad\qquad
U = (1+w_d)v_m - u_d\,.
\label{deu}
\ee
It can be seen that the adiabaticity between dark energy and matter perturbations corresponds to $\De = U = 0$.
Using (\ref{ded}) - (\ref{vm})
The evolution equations for $\De$ and $U$ can be written as
\ba
\fl
\De' &=& \h r \De - k U\,,
\label{de1}\\
\fl
U' &=& \h\(3w_d - 1\)U + \h w_d \(\frac 2\al - 4\) v_m - \frac k3 \(\frac 2\al - 1\) \De + w_d\frac k3  \(\frac 2\al - 4\) \de_m
+ (1 + w_d) \frac{\rho_b}{\rho_m} \Tc\,.
\nonumber\\
\fl
\label{du1}
\ea
From these equations, one can show that on superhorizon scales the evolution of $\De$ is governed by
\ba
\De'' &+& \h\(1 - 3w_d - r\)\De' - \(\h' + \h^2\)r\De 
\nonumber\\
&\simeq& - \h w_d \(\frac 2\al - 4\) k v_m
- w_d\frac{k^2}3  \(\frac 2\al - 4\) \de_m
- (1 + w_d) \frac{\rho_b}{\rho_m} k \Tc\,,
\label{ddde1}
\ea
and on subhorizon scales $\De$ obeys
\ba
\De'' &+& \h\(1 - 3w_d - r\)\De' - \frac{k^2}3 \(\frac 2\al - 1\) \De
\nonumber\\
&\simeq&
-\h w_d \(\frac 2\al - 4\) k v_m - w_d\frac{k^2}3  \(\frac 2\al - 4\) \de_m
- (1 + w_d) \frac{\rho_b}{\rho_m} k \Tc\,.
\label{ddde2}
\ea
Here, we keep the terms on the RHS of (\ref{ddde1}), because 
in the case of our interest $\De$ can be very small so that its evolution on large scales can be affected by these terms.

To study how $\De$ evolves, we start with the simplest case where 
the main contributions to the matter perturbations in (\ref{dp}) come from CDM perturbations,
i.e., $\de_m \approx \de_c$ and $v_m \approx v_c$, and
the contribution from $\Tc$ becomes negligible.
In the radiation and matter eras, $w_d \approx 0$, so that the second and fourth terms on the RHS of (\ref{du1}) can be neglected.
Hence, it follows from (\ref{de1}) and (\ref{du1}) that if the perturbations in dark energy and matter are initially adiabatic in the early epoch,
i.e., $\De = U =0$, they will remain adiabatic at later time as long as $w_d \approx 0$
both on superhorizon and subhorizon scales.
Nevertheless, if the perturbations in dark energy and matter are not adiabatic in the early epoch,
the non-adiabatic perturbations can grow both inside and outside the horizon.
During the radiation and matter eras, the RHS of (\ref{ddde1}) can be neglected,
so that the solutions of (\ref{ddde1}) are
\be
\De = \left\{
\begin{array}{ll}
c_1 \eta^r + c_2 & \mbox{ during radiation domination}, \\
c_3 \eta^{2r}  + c_4 \eta^{-1} & \mbox{during matter domination}
\end{array}\right. \,.
\ee
It can be seen that $\De$ has a growing mode, and the constant adiabatic mode corresponds to $c_1 = c_2 = c_3 = c_4 =0$.
Similarly, during the radiation and matter eras, (\ref{ddde2}) becomes a homogeneous differential equation.
To solve this equation, we use the variable $\chi = \De / \eta^p$, where $p = (r -1)/2$ and $p = r-1$
during radiation and matter domination respectively.
In terms of this variable, (\ref{ddde2}) becomes
\be
\chi'' - \[\frac{k^2 r_{\rm early}}{3} + \frac 1{\eta^2}\(\mu^2 - \frac 14\)\]\chi \simeq 0\,,
\ee
where $\mu = r_{\rm early} /2$ and $\mu = r_{\rm early} - 1/2$ during radiation and matter eras respectively.
Since $k\eta \gg 1$ inside the horizon, the non-adiabatic perturbations grow inside the horizon as
\be
\De = \left\{
\begin{array}{ll}
d_1 \eta^{(r - 1)/2} {\rm e}^{\sqrt{r_{\rm early}/3}\, k\eta} 
+d_2 \eta^{(r - 1)/2} {\rm e}^{-\sqrt{r_{\rm early}/3}\, k\eta}
& \mbox{during radiation domination,}\\
d_3 \eta^{r - 1} {\rm e}^{\sqrt{r_{\rm early}/3}\, k \eta} 
+ d_4 \eta^{r - 1} {\rm e}^{-\sqrt{r_{\rm early}/3}\, k \eta} 
& \mbox{during matter domination,}\\
\end{array}\right. \,.
\ee
Again, the constant adiabatic mode corresponds to $d_1=d_2=d_3=d_4=0$, and $|\De |$ can grow inside the horizon.
Since $1/3 < \al < 1$, $r$ is in the range $1 < r < 5$ during the radiation and matter eras.
Thus, from small non-adiabatic perturbations between dark energy and matter in the early epoch,
$|\De |$ and also $|U|$ can grow by many orders of magnitude, and can become unphysically large at late time.
This implies an unphysically large amplitude of $\de_d$ and also $\de_m$ together with the metric perturbation
due to the back reaction which follows from the perturbed einstein equation \cite{Doran:03},
\ba
3H\(\dot\Phi + H \Psi\) + \frac{k^2}{a^2}\Phi 
&=& - \frac{3H^2}2 \(\Om_r\de_r + \Om_m\de_m + \Om_d\de_d\) 
\nonumber\\
&=& -\frac{3H^2}{2}\(\Om_r\de_r + (\Om_m + \Om_d)\de_m - \Om_d\De \)\,,
\label{e00}
\ea
where a subscript $r$ denotes radiation including photon and neutrino,
and a subscript $m$ denotes matter including CDM and also baryon in general case.

\begin{figure}[ht]
\includegraphics[height=0.4\textwidth, width=0.9\textwidth,angle=0]{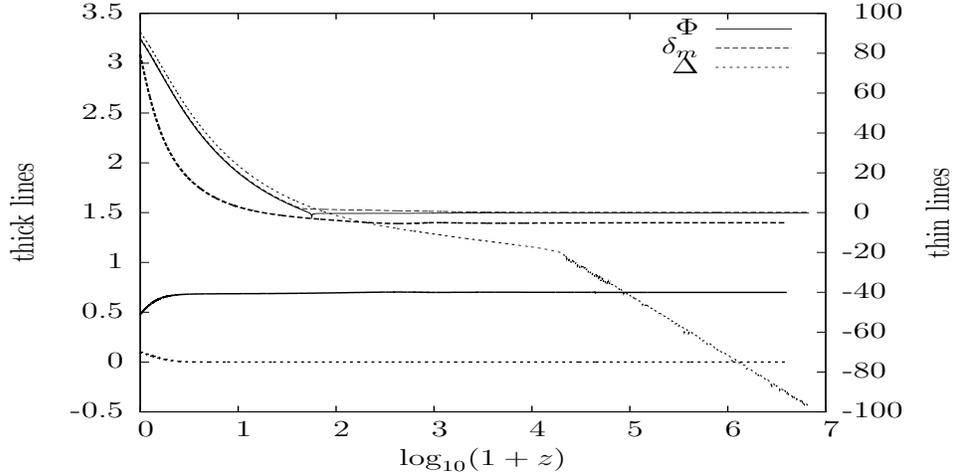}
\caption{
The evolution of $\Phi$, $\de_m$ and $\De$.
The thick lines represent the perturbations mode that enters the horizon about the present epoch, $k = 2.4\times 10^{-4}$Mpc$^{-1}$,
while the thin lines represent the perturbations mode that enters the horizon about the time of radiation-matter equality, $k = 1.6\times 10^{-2}$Mpc$^{-1}$ .
For the thin lines, we plot $\log_{10}|\Phi|$, $\log_{10}|\de_m|$ and $\log_{10}|\De|$ instead of their actual values.
}
\label{fig:2}
\end{figure}

\begin{figure}[ht]
\includegraphics[height=0.4\textwidth, width=0.9\textwidth,angle=0]{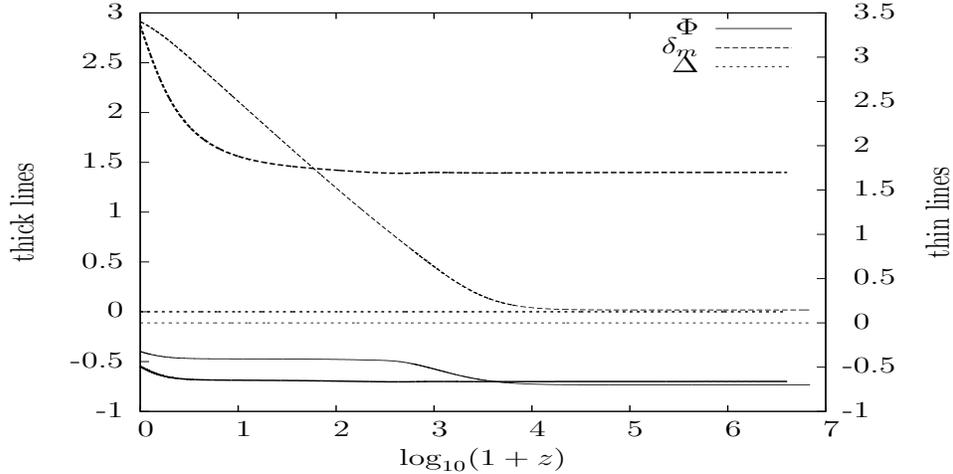}
\caption{
The evolution of $\Phi$, $\de_m$ and $\De$.
The thick lines represent the perturbations mode that enters the horizon about the present epoch, $k = 2.4\times 10^{-4}$Mpc$^{-1}$,
while the thin lines represent the perturbations mode that enters the horizon about the time of radiation-matter equality, $k = 1.6\times 10^{-2}$Mpc$^{-1}$ .
For the thin lines, we plot $\log_{10}\de_m$ instead of $\de_m$.
}
\label{fig:3}
\end{figure}

Roughly speaking, the existence of the unphysically large density perturbations, which implies an instability problem in dark energy,
can be avoided if the non-adiabatic perturbations between dark energy and matter disappear or have a very small amplitude.
However, near the end of matter domination, these non-adiabatic perturbations can be generated although they disappear in the early epoch.
This is because $w_d$ becomes negative, so that (\ref{ddde1}) and (\ref{ddde2}) become inhomogeneous.
Obviously, the constant adiabatic solution $\De = U =0$ cannot be the particular solutions of these equations.
On large scales, the terms on the RHS of (\ref{ddde1}) cannot influence the evolution of $\De$ so much 
because their amplitude is small.
Nevertheless, inside the horizon, to create the structure in the universe, $\de_m$ must grow, consequently
the inhomogeneous part of (\ref{ddde2}) leads to rapid growth of $|\De|$.
As a result, $\de_d$, $\de_m$ and the metric perturbation rapidly grow and become unphysically large.
However, it is easy to see that the inhomogeneous part of (\ref{ddde2}) will always vanish if $\al = 1/2$,
and hence the generation of non-adiabatic mode after matter domination can be avoided for this value of $\al$.
These qualitative considerations are in agreement with the numerical integration presented in figures (\ref{fig:2}) and (\ref{fig:3})
for the case where $\al = 0.4$ and $\al = 1/2$ respectively.
In these figure, the initial conditions for $\De$ and $U$ are chosen such that $\De = U =0$,
and the perturbation variables are normalized such that the curvature perturbation $\zeta$ defined below in (\ref{zeta}) equals to 4.
The numerical solution of the evolution equations for the perturbations and also the CMB power spectrum are obtained using CMBEASY \cite{cmbeasy}.
From these figures one can see that the instability problem in Ricci dark energy can be avoided
if $\al = 1/2$.

It follows from the above consideration that if the contributions from baryon perturbations in (\ref{dp}) can be neglected,
the instability problem in Ricci dark energy can be avoided if the perturbations in dark energy and matter are adiabatic initially, and $\al = 1/2$.
Let us now compute the initial conditions for the perturbations in dark energy and the other species
needed for the numerical integration for density perturbations and CMB power spectrum.
Since the energy density of Ricci dark energy depends on Ricci scalar,
the amplitude of the density perturbation in this dark energy depends on the metric perturbations.
The relation between the density and metric perturbations can be obtained from (\ref{rc}) as \cite{Feng:09, Riotto:02}
\be
\de\rho_d = \frac{\alpha}{2} \de R = \al \bigg[2\frac{k^2}{a^2} \Phi - 4 \frac{k^2}{a^2} \Psi - 3\ddot\Phi - 6H\dot\Phi - 9H\dot\Psi -6(\dot H+ 2H^2)\Phi
    \bigg]\,,
\ee
where the vanishing of the anisotropic perturbations in radiation is not assumed.
Hence, on the superhorizon scales and in the radiation dominated epoch,
\be
\fl
\de_d = \frac{\de\rho_d}{\rho_d} = - \frac{12}{\rho_m +\rho_d}\(\dot H+ 2H^2\)\Phi = -\frac{4}{\Om_m + \Om_d}\(- 2 \Om_r - \frac 32 \Om_m - \frac 32 \Om_d + 2\)\Phi = - 2\Phi\,.
\label{de-phi}
\ee
To obtain the above equation, we suppose that $\Phi$ is nearly constant on large scales
because we use the adiabatic initial conditions for photon and neutrino, and the adiabatic perturbations between dark energy and matter $\De = 0$  is the constant mode.
Using the adiabatic initial conditions derived in \cite{Doran:03}, the relation between the initial value of the density contrast of photon $\de_\gamma$ and metric perturbations
can be written in Newtonian gauge as
\be
\de_\gamma = - \frac{10}{5 + 2\Om_\nu}\Phi\,.
\ee
This implies that in the early epoch if the perturbations in matter and dark energy are adiabatic, i.e., $\de_d = \de_m = -2\Phi$,
the perturbations in matter and photon cannot be adiabatic.
This conclusion holds even though the anisotropic perturbations in neutrino are neglected,
because, in this case, $\de_r = - 2\Phi = \de_m$ in the early epoch.

\begin{figure}[ht]
\includegraphics[height=0.4\textwidth, width=0.9\textwidth,angle=0]{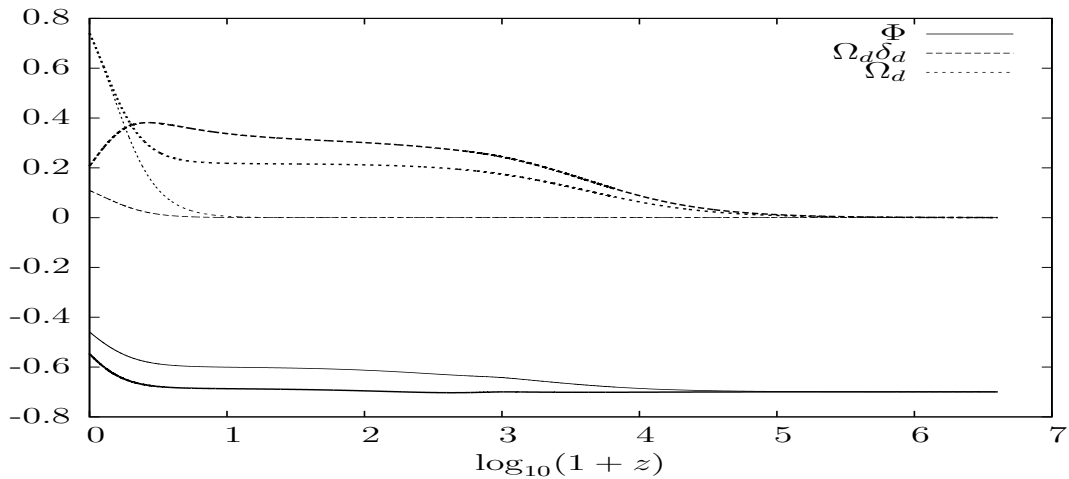}
\caption{
The large scale evolution of $\Phi$, $\de_d$ and $\Om_d\de_d$
for Ricci dark energy model (thick lines)
and dark energy model with constant equation of state parameter $w_d = -0.9$ (thin lines).
For later model, we use adiabatic initial conditions for all species,
while forRicci dark energy model we set $\al = 1/2$ and $\De = U = 0$ initially.
}
\label{fig:4}
\end{figure}

To study how the non-adiabaticity between radiation and matter perturbations influences OSW effect,
we derive the evolution equation for the curvature perturbations on slices of uniform total energy density by ignoring the anisotropic perturbations in photon and neutrino ,
and treating photon and neutrino as a single perfect fluid.
Furthermore, since in the early epoch dark energy evolves like matter and $\de_d = \de_m$,
we also treat dark energy and matter as a single fluid  whose density contrast$\de_{\tilde m}$  obeys (\ref{dem}).
Here, a subscript $\tilde{m}$ denotes this combined fluid.
The evolution of the density contrast of the total energy density $\de_T = \de\rho_T / \rho_T$,
where $\de\rho_T = \de\rho_r + \de\rho_{\tilde m}$ and $\rho_T = \rho_r + \rho_{\tilde m}$,
is governed by the general evolution equation for the perfect fluid
\be
\de_T' = - 3\h\(\frac{\de p_T}{\de\rho_T} - w_T\)\de_T + 3\(1+w_T\)\Phi' - k (1+w_T)v_T\,,
\label{det}
\ee
where $w_T = (1/3)\rho_r / \rho_T$ is the total equation of state parameter,
and $v_T = v_r + v_{\tilde m}$ is the total velocity perturbations.
From the definition of the curvature perturbations on slices of uniform total energy density \cite{Doran:03}
\be
\zeta = - \Phi + \frac{\de_T}{3(1+w_T)}\,,
\label{zeta}
\ee
and (\ref{det}), one can show that on large scales
\be
\zeta' = - \frac{\h}{\rho_T + p_T}\(\de p_T - c_a^2\de\rho_T\) = - \frac{\h}{\rho_T + p_T}p_{\rm nad}\,,
\ee
where $c_a^2 = (1/3)\rho_r' / \rho_T'$ is the total adiabatic sound speed.
It follows from the above equation that if the perturbation in radiation and fluid $\tilde{m}$ are adiabatic, $p_{\rm nad}$ will vanish. Consequently, $\zeta$ is constant on large scales (cee e.g. \cite{Wands:00}).
In the radiation dominated epoch, the main contribution to $\zeta$ comes from the curvature perturbation on the hypersurfaces of uniform radiation energy density,
$\zeta_{\rm rd} = -\Phi_{\rm rd} + \de_{r\, {\rm rd}} / 4 = - 3\Phi_{\rm rd} /2$,
where the subscript rd denotes the value in radiation dominated epoch.
Similarly, during the matter domination, $\zeta_{\rm md} = -\Phi_{\rm md} + \de_{\tilde{m}\, {\rm md}}/3 = - 5\Phi_{\rm md} /3$,
where the subscript md denotes the value in the matter dominated epoch.
Thus, the constancy of $\zeta$ yields $\Phi_{\rm md} = 9\Phi_{\rm rd}/10$ \cite{Hu:95}.
In the case of non-adiabatic perturbation, we have to know how the entropy perturbations in radiation and ${\tilde{m}}$ fluid evolve.
According to the conservation of energy, $\de_r$ obeys
\be
\de_r' = 4\Phi' - \frac 43 k v_r\,.
\label{der}
\ee
Using the above equation and (\ref{dem}), we get $S' = - k(v_r - v_{\tilde m})$, where
$S = 3\de_r/4 - \de_{\tilde{m}}$ is the entropy perturbations between radiation and $\tilde{m}$ fluid.
Hence, on large scales $S \simeq$  constant \cite{Hu:95}, so that $\de_r \simeq \de_{\rm m}$ until the matter domination.
However, the detail analysis shows that the non-adiabatic perturbations between radiation and matter can grow in early epoch \cite{Brown:11}.
Since this growth of  non-adiabatic perturbations between radiation and matter does not influence our rough estimation below,
we assume that $S \simeq$ constant, and therefore
\be
p_{\rm nad} = \de p_r - \frac 13 \frac{\rho_r'}{\rho_T'}\frac{\rho_T}{\rho_r}\de\rho_r= \frac 13 \de \rho_r - \frac 13\frac{4}{4\Om_r + 3\Om_{\tilde{m}}}\de\rho_r\,,
\ee
which vanishes in the radiation dominated epoch and then becomes negative when $\Om_r < 1$,
because in our notations $\Phi$ is negative and $\de\rho_r$ is positive.
This implies that $\zeta$ increases during the transition from radiation domination to matter domination,
so that $\zeta_{\rm rd} < \zeta_{\rm md}$ and hence $|\Phi_{\rm md}| > 9|\Phi_{\rm rd}| / 10$,
which is in agreement  with the numerical solution plotted in figure (\ref{fig:4}).
Using the approximations $\de_r \approx \de_{\tilde m} \simeq - 2\Phi$ and $\Psi \simeq \Phi$ on large scales during last scattering epoch, one gets
\be
\Theta_0 + \Psi \simeq \Theta_0 + \Phi = \frac 14 \de_r + \Phi = \frac 12 \Phi\,,
\ee
where $\Theta_0$ is the temperature perturbations in Newtonian gauge \cite{Hu:95} and the above expression is evaluated at last scattering.
This equation shows that the non-adiabaticity between radiation and matter perturbations enhances the OSW effect
compared with the adiabatic case in which $\Theta_0 + \Psi = \Psi /3$.

\begin{figure}[ht]
\includegraphics[height=0.4\textwidth, width=0.9\textwidth,angle=0]{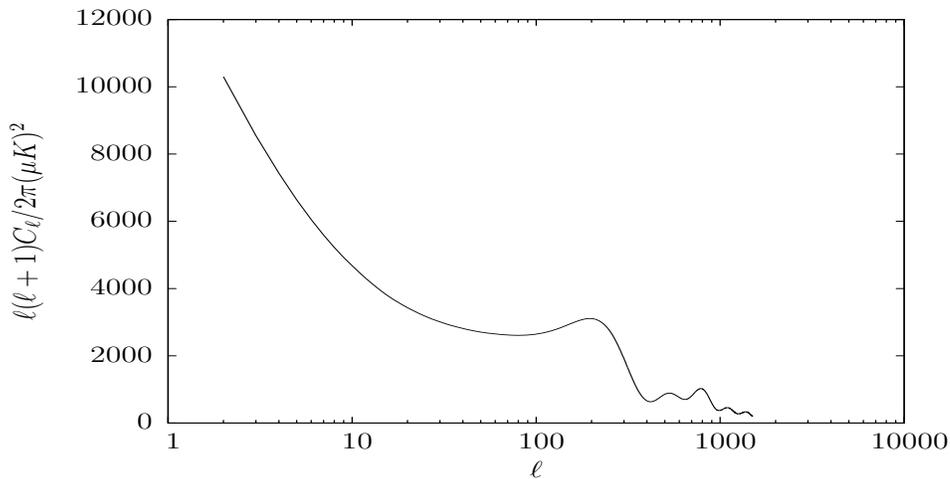}
\caption{
The CMB power spectrum for the case where $\al = 1/2$ and $\De = U = 0$ initially.
}
\label{fig:5}
\end{figure}

The main contributions from the density perturbations in Ricci dark energy to the density perturbations in the universe come
from the ISW effect which may be represented in terms of $\dot\Phi$.
On large scales and after last scattering, $\dot\Phi$ is governed by the perturbed Einstein equation
\be
\dot\Phi = - H \Phi - \frac{H}2 \(\Om_m\de_m + \Om_d\de_d\)\,,
\ee
Here, we have supposed that $\Phi \approx \Psi$.
Since $\Phi$ and the density contrast have opposite signs in our cases,
The contribution from $\de_d$ can reduce $|\dot\Phi|$ because it cancels the contribution from the term $- H \Phi$.
The contribution from $\de_d$ increases if $\Om_d$ during matter era increases,
and when the contribution from $\de_d$ increases $|\dot\Phi|$ decreases.
Usually OSW effect is partially canceled by ISW effect \cite{Hu:95}, a smaller $|\dot\Phi|$ leads to a smaller cancellation
and therefore a large amplitude of CMB power spectrum at low multipoles.
For the case of Ricci dark energy, $\Om_d$ is not small during the matter era, so that the contribution from $\de_d$
can significantly reduce $|\dot\Phi|$, leading to a large amplitude of the CMB spectrum at low multipoles.
 The reduction of $|\dot\Phi|$ due to the contribution from $\de_d$ is shown in figure (\ref{fig:4}).
In figure (\ref{fig:5}), the CMB power spectrum for the case
where the instability problem in Ricci dark energy can be avoided, i.e.,  $\De = U =0$ initially and $\al = 1/2$, is plotted.
It can be seen from the figure that the CMB power spectrum is greatly enhanced at small multipoles because the SW effect is modified due to the density perturbations in Ricci dark energy.

\begin{figure}[ht]
\includegraphics[height=0.4\textwidth, width=0.9\textwidth,angle=0]{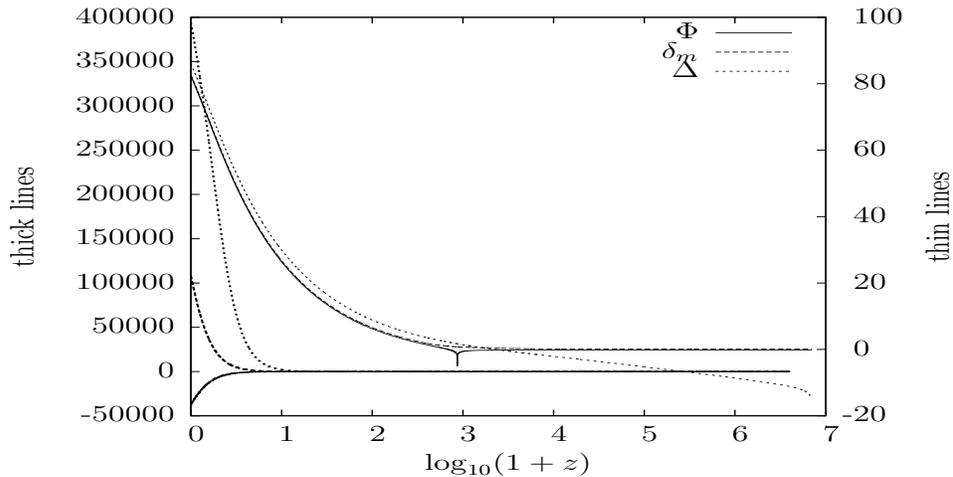}
\caption{
The evolution of $\Phi$, $\de_m$ and $\De$.
The thick lines represent the perturbations mode that enters the horizon about the present epoch, $k = 2.4\times 10^{-4}$Mpc$^{-1}$,
while the thin lines represent the perturbations mode that enters the horizon about the time of radiation-matter equality, $k = 1.6\times 10^{-2}$Mpc$^{-1}$ .
For the thin lines, we plot $\log_{10}|\Phi|$, $\log_{10}|\de_m|$ and $\log_{10}|\De|$ instead of their actual values.
}
\label{fig:6}
\end{figure}

Finally, we include the baryon perturbations in $\de\rho_m$ in (\ref{dp}),
so that the term $\Tc$ on the RHS of (\ref{du1}) cannot be neglected.
Using the same consideration as above, one can see that this term can generate a growing non-adiabatic mode
inside the horizon, leading to an instability problem in Ricci dark energy model.
Up to our trial, the generation of this non-adiabatic mode cannot be avoided by adjusting the initial conditions for the perturbations, special values of $\al$
or can be canceled by the other terms, hence the instability problem is not easy to avoid in this case.
The evolution of $\Phi$, $\de_m$ and $\De$ for this case is plotted in figure (\ref{fig:6}).

 \section{Conclusions}

We study the evolution of density perturbations in Ricci dark energy model.
We extend the analysis in \cite{Feng:09} by allowing the non-adiabatic perturbations between matter and dark energy.
In the case where the contributions from baryon perturbations are neglected,
if the perturbations in matter and dark energy are adiabatic initially,
they will remain adiabatic both on superhorizon and subhorizon scales until the equation of state parameter of dark energy becomes negative.
When the equation of state parameter of dark energy becomes negative, the non-adiabatic perturbations between matter and dark energy perturbations will be generated even though
they disappear in the early epoch. This non-adiabatic mode is a growing mode
that can lead to an instability in Ricci dark energy, i.e., the density perturbations in dark energy, matter, and metric perturbations grow extremely large.
The generation of this non-adiabatic mode can be avoided if $\al = 1/2$ and the perturbations in dark energy and matter
are adiabatic initially.
Nevertheless, the perturbations in matter and photon cannot be adiabatic in the early epoch, so that
the OSW effect is modified.
The main influence of the perturbations in dark energy on the density perturbations in the universe is
the great modification of the ISW effect due to a large amount of dark energy during matter era.
This modification of ISW effect greatly enhances the CMB spectrum at low multipoles.
When the contributions from baryon perturbations are included, the coupling between baryon and photon
can generate growing non-adiabatic mode that leads to the instability in Ricci dark energy.
In this case, the generation of growing non-adiabatic mode cannot be avoided by simply adjusting the initial conditions for the density perturbation or spacial choices of $\al$.

\section*{Acknowledgments}

The authors would like to thank A. Ungkitchanukit
for comments on the manuscript.
This work is supported by Thailand Research Fund (TRF) through grant RSA5480009.

\end{document}